\documentclass[aps, prl, showpacs, twocolumn, amsfonts, amsmath, amssymb, superscriptaddress, floatfix]{revtex4-1}

\usepackage[T1]{fontenc}
\usepackage[latin9]{inputenc}
\usepackage{color}
\usepackage{graphicx}
\usepackage{ulem}
\usepackage{filemod}

\def\be{\begin{equation}}
\def\ee{\end{equation}}
\def\bea{\begin{eqnarray}}
\def\eea{\end{eqnarray}}

\begin{document}

\title{Three-dimensional localized-delocalized Anderson transition in the time domain}


\author{Dominique Delande}
\affiliation{Laboratoire Kastler Brossel, UPMC-Sorbonne Universit\'es, CNRS, ENS-PSL Research University, Coll\`ege de France, 4 Place Jussieu, 75005 Paris, France}

\author{Luis Morales-Molina}
\affiliation{Departamento de F\'isica, Facultad de F\'isica,
 Pontificia Universidad Cat\'olica de Chile, Casilla 306, Santiago 22, Chile}

\author{Krzysztof Sacha} 
\affiliation{
Instytut Fizyki imienia Mariana Smoluchowskiego, 
Uniwersytet Jagiello\'nski, ulica Profesora Stanis\l{}awa \L{}ojasiewicza 11, 30-348 Krak\'ow, Poland}
\affiliation{Mark Kac Complex Systems Research Center, Uniwersytet Jagiello\'nski, ulica Profesora Stanis\l{}awa \L{}ojasiewicza 11, 30-348 Krak\'ow, Poland
}


\begin{abstract}
Systems which can spontaneously reveal periodic evolution are dubbed time crystals. This is in analogy with space crystals that display periodic behavior in  configuration space. While space crystals are modeled with the help of space periodic potentials, crystalline phenomena in time can be modeled by periodically driven systems. Disorder in the periodic driving can lead to Anderson localization in time: the probability for detecting a system at a fixed point of configuration space becomes exponentially localized around a certain moment in time. We here show that a three-dimensional system exposed to a properly disordered pseudo-periodic driving may display a localized-delocalized Anderson transition in the time domain, in strong analogy with the usual three-dimensional Anderson transition in disordered systems. Such a transition could be experimentally observed with ultra-cold atomic gases.
\end{abstract}

\maketitle

The fact that a detector, placed at some position, has a large probability to click at a certain moment in time when a particle is passing nearby is not very surprising. It is more interesting when this localization has universal characteristics such as an exponential
	shape. This is the case when temporal Anderson localization is induced by a fluctuating driving force. 

Usual Anderson localization is the configuration space exponential localization of eigenstates in the presence of a spatially disordered potential \cite{Anderson1958}. 
It is accompanied by the inhibition of transport due to destructive interference between different multiple scattering paths. Anderson localization may also take place in momentum space  -- where it is called ''dynamical localization'' -- e.g. in the so-called kicked rotor, where an effective pseudo-disorder is induced by the classically chaotic dynamics \cite{Fishman:LocDynAnders:PRL82,Moore:AtomOpticsRealizationQKR:PRL95,Casati:IncommFreqsQKR:PRL89}. 

The interference between paths scattered by a disorder depends on their geometrical properties and especially on the dimension of a system. According to the scaling theory of localization \cite{Abrahams:Scaling:PRL79}, one-dimensional (1D) and time-reversal invariant spinless 2D systems reveal localization regardless how weak the disorder is. In the 3D case, the situation is more complicated since the scaling theory predicts a second order phase transition --- for a fixed disorder strength, all eigenstates of a system with energies up to the so-called mobility edge are localized and the other ones are not. How the mobility edge depends on the disorder 
has been analyzed in a variety of systems \cite{MuellerDelande:Houches:2009}. A rotor driven by a quasi-periodic sequence of kicks with $d$ quasi-periods can be mapped
on a $d$-dimensional pseudo-disordered system, allowing for a simple experimental method
for studying Anderson localization in dimension $d.$ This made it possible to investigate theoretically and experimentally the critical behavior in the vicinity of the 3D mobility edge \cite{Lemarie:Anderson3D:PRA09,Manai:Anderson2DKR:PRL15}.

Anderson localization in the time domain can be realized in systems that are perturbed by a time fluctuating force, provided the latter is repeated periodically with a frequency that is resonant with the unperturbed motion of the system \cite{Sacha2015a,sacha_delande16}. In classical mechanics, the fluctuating force produces a diffusive motion. In the quantum description, interference effects cause the system to localize. That is, if we put a detector close to any point of the trajectory, we will observe that the detection probability is exponentially localized around a certain moment in time. Moreover, this exponential profile comes back every period of the classical motion. Thus, we deal with a situation analogous to Anderson localization of a particle on a ring (periodic boundary conditions) in the presence of a disordered time-independent potential. By traveling periodically around the ring, one observes periodically a localized density profile. 

Anderson localization in time belongs to more general phenomena dubbed time crystals \cite{Wilczek2012,Li2012}. Time crystals are systems that can spontaneously switch to a periodic motion. That is, even if they are prepared in an eigenstate, which possesses continuous time translational symmetry, a small perturbation can push them to periodic motion. There is a debate in the literature whether continuous time translational symmetry can be spontaneously broken \cite{Chernodub2013,Wilczek2013,Bruno2013,Wilczek2013a,Bruno2013a,Li2012a,Bruno2013b,Watanabe2015}. So far it has been shown \cite{Sacha2015,khemani16,else16,Keyserlingk16,yao17,Weidinger16}, and demonstrated experimentally \cite{zhang16,choi16}, that spontaneous breaking of a discrete time translational symmetry to another discrete one is possible. Here, we will not consider this problem of spontaneous formation of time crystals. We will model crystalline behavior by periodically driven systems \cite{Guo2013,Sacha2015a,sacha_delande16}, in analogy with condensed matter physics where spatially periodic potentials are used to model space crystals. 

Time is a single degree of freedom, therefore, we cannot expect multidimensional time crystals. In this Letter, we show that time crystal phenomena with properties of multidimensional condensed matter systems can be observed. More precisely we demonstrate the analog in the time domain of the usual three-dimensional localized-delocalized Anderson transition \cite{MuellerDelande:Houches:2009}. 

Let us consider a particle with a unit mass moving in the 3D space with periodic boundary conditions (3D torus) whose position is denoted by three angles: $\theta$, $\psi$ and $\phi$. We assume that the particle is perturbed by a temporally disordered potential, i.e. the Hamiltonian of the system reads
\be
H=\frac{p_\theta^2+p_\psi^2+p_\phi^2}{2}+V_0g(\theta)g(\psi)g(\phi)f_1(t)f_2(t)f_3(t), 
\label{h3d}
\ee
where $V_0$ is the amplitude of the perturbation. The time dependent functions are periodic but between $t$ and $t+2\pi/\omega_i$ they perform random fluctuations, i.e. $f_i(t+2\pi/\omega_i)=f_i(t)=\sum_{k\ne 0}f_k^{(i)}e^{ik\omega_i t}$ where $f^{(i)}_k=f^{(i)*}_{-k}$ are independent random numbers. We assume that the ratios of the frequencies $\omega_i$ are irrational numbers. In contrast, $g(x)$ is assumed to be a regular function --- we choose $g(x)=x/\pi$ for $x\in[-\pi,\pi[$, i.e., $g(x)=\sum_{n}g_ne^{inx}$ where $g_n\!=\!\frac{i(-1)^{n}}{\pi n}$ for $n\ne 0$ and $g_0\!=\!0$. 
Thus, we deal with a perturbation which behaves regularly in the configuration space (for fixed time) but which is disordered in time. As shown below, in order to observe Anderson localization in the time domain, it is important
that both the spatial function $g(x)$ and the temporal disorder $f_i(t)$ contains many Fourier components.
For the sake of simplicity, we
choose the $f_k^{(i)}$ components so that:
\begin{equation}
|g_kf_k^{(i)}|=\frac{1}{\sqrt{k_0}\pi^{1/4}}e^{-k^2/(2k_0^2)}
\end{equation} 
with $k_0$ a free-to-choose parameter and Arg$(f_k^{(i)})$ (for $k>0$) are independent random variables chosen uniformly in the interval $[0,2\pi[.$  Such a Gaussian shape makes the computation of the localization length easy in 1D~\cite{sacha_delande16}, but any similar shape will lead to a 3D metal-insulator Anderson transition in the time domain. This is a robust phenomenon that takes place in 3D as shown below. Similarly, any form of $g(x)$ with sufficiently many $k$ components can be used.

Let us switch to the moving frame where $\Theta=\theta-\omega_1 t$, $\Psi=\psi-\omega_2 t$ and $\Phi=\phi-\omega_3 t$. In this frame, $\Theta$, $\Psi$ and $\Phi$ are slowly varying variables if we choose the conjugate momenta $P_\Theta=p_\theta-\omega_1\approx 0$,  $P_\Psi=p_\psi-\omega_2\approx 0$ and $P_\Phi=p_\phi-\omega_3\approx 0$. In the secular approximation \cite{Buchleitner2002,Lichtenberg_s}, the dynamics of the slowly varying variables is described by an effective Hamiltonian obtained by averaging the original Hamiltonian over time~\cite{supplement}:
\be
H_{\mathrm{eff}}=\frac{P_{\Theta}^2+P_{\Psi}^2+P_\Phi^2}{2}+V_{\mathrm{eff}}(\Theta,\Psi,\Phi),
\label{heff}
\ee
with $V_{\mathrm{eff}}=V_0h_1(\Theta)h_2(\Psi)h_3(\Phi)$
where 
\be
h_i(x)=\sum_{k\ne 0}g_k f_{-k}^{(i)}e^{ikx}.
\label{Eq:defh}
\ee
In Eq. (\ref{heff}), the constant term $(\omega^2_1+\omega^2_2+\omega^2_3)/2$ is omitted. To obtain $H_{\mathrm{eff}}$ we take advantage of the fact that the ratios of the frequencies $\omega_i$ are irrational numbers. The first order of the secular approximation (\ref{heff}) is valid provided the amplitude  of the perturbation $V_0$ is small or $\omega_i$'s are large and fulfill the relations $\omega_1>2k_0(\omega_2+\omega_3)$, i.e. the second order correction terms, $V_0^2/(\sum_in_i\omega_i)^2e^{-\sum_in_i^2/4k_0^2}$, never suffer from a small denominator problem and are negligible~\cite{supplement}.

The effective potential $V_{\mathrm{eff}}(\Theta,\Psi,\Phi)$ is a product of three independent disordered potentials $h_i$ along each degree of freedom. It can be characterized by its two-point
correlation function which is trivially factorized as a product of three times the same correlation function along the three directions. In the limit of relatively large $k_0$ we are interested in, there is a large number $\simeq k_0$ of random contributions in the sum (\ref{Eq:defh}); from the central limit theorem, we deduce that $h_i(x)$ has a Gaussian distribution with zero mean. The correlation function is easily computed and, for large $k_0$, it reads 
\begin{equation}
\overline{h_i(x')h_i(x'+x)} = V_0^2 \exp \left( - \frac{x^2}{2\sigma^2}\right),
\end{equation}   
where  $\overline{\phantom{X}}$ denotes the averaging over disorder realizations.
$\sigma=\sqrt{2}/k_0$ is the correlation length of the disordered potential.  

Let us assume, for a moment, that $\Theta$, $\Phi$ and $\Psi$ are not limited to the interval $[0,2\pi[$ but extend from minus infinity to infinity.
$V_{\mathrm{eff}}(\Theta,\Psi,\Phi)$ being a generic 3D random potential, one expects a
localized-delocalized transition to take place at some value of the energy $E_c$ called the mobility edge. There are three different energy scales in the problem: the strength $V_0$ of the potential,
the energy $E$ of the particle and the so-called correlation energy
\begin{equation}
E_\sigma = \frac{1}{\sigma^2} = \frac{k_0^2}{2}.
\end{equation}
$E_\sigma$ sets the natural energy scale of the problem~\cite{Kuhn:Speckle:NJP07}, so that the ratio $E_c/E_\sigma$ depends only on the ratio $V_0/E_\sigma.$
 
\begin{figure}
	\includegraphics[width=0.8\columnwidth]{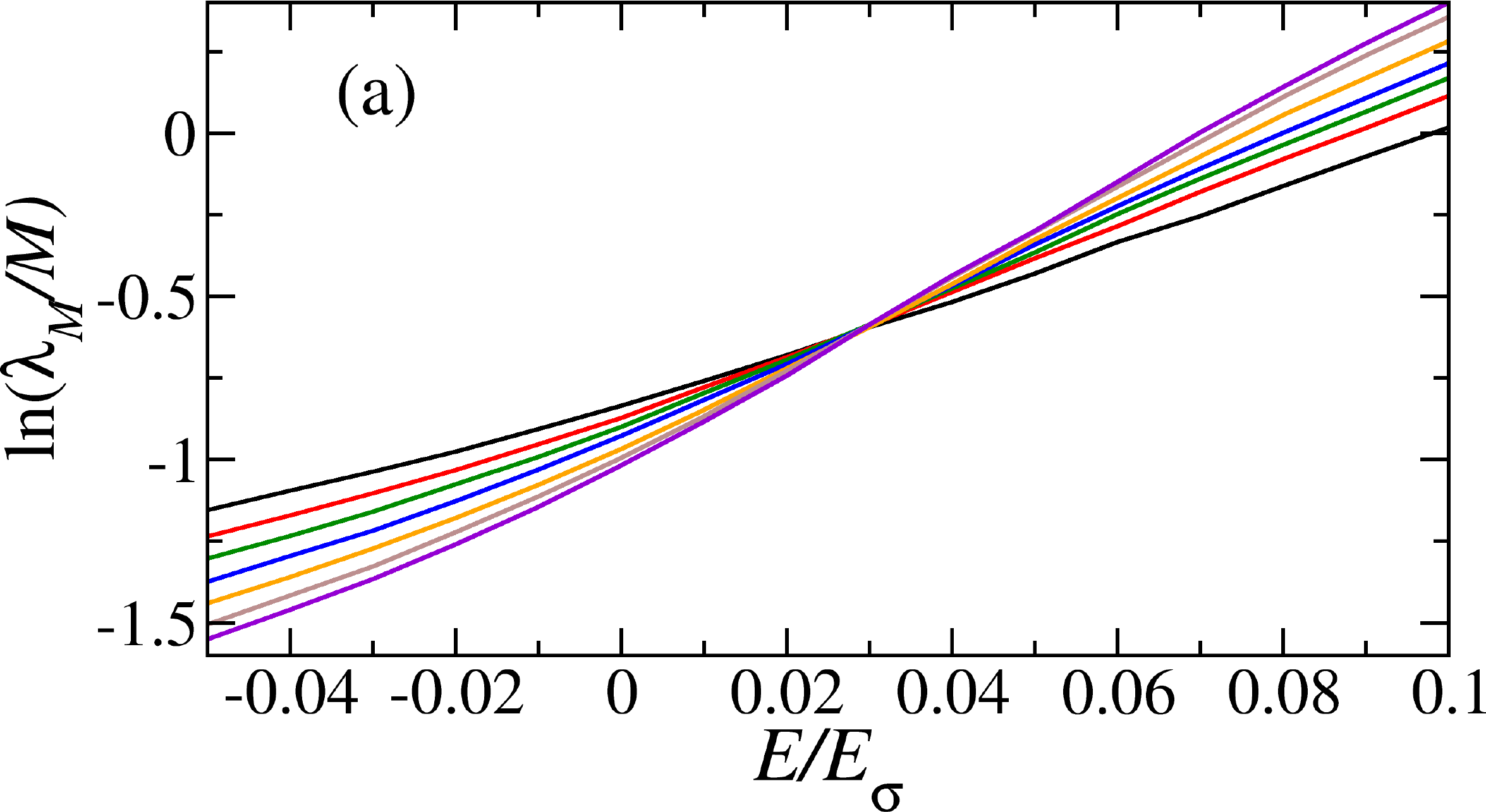}
	\vskip 0.2cm
	\includegraphics[width=0.75\columnwidth]{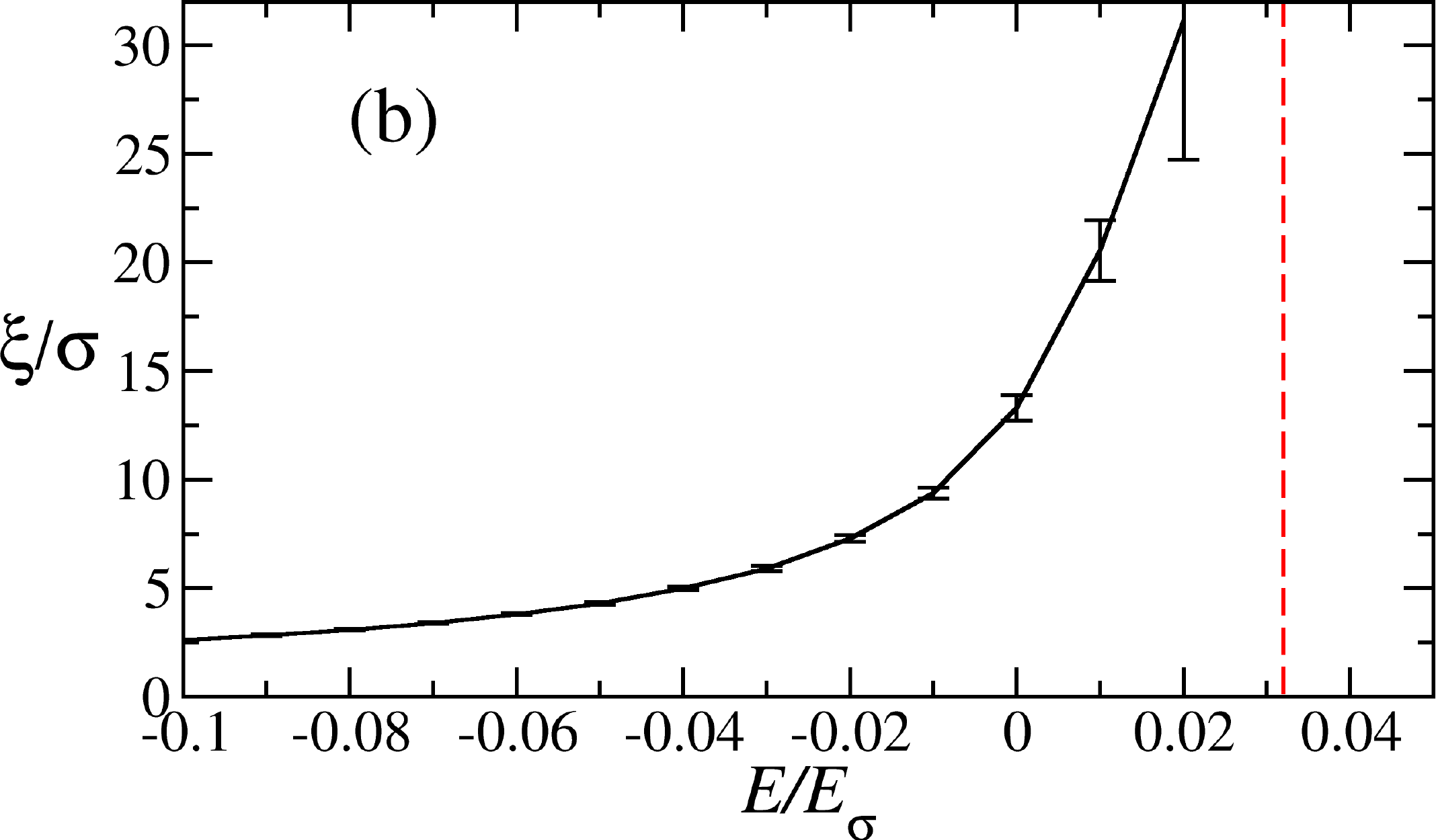}
\caption{Numerical determination of the mobility edge for $V_0\!=\!E_\sigma.$ For a given energy, we compute the localization length
$\lambda_M$ of a long bar-shaped grid with square section $M\!\times\!M$. 
(a): Each curve is computed for a single $M$ value from 25 to 55 with step 5 (slope increases with $M$) and shows $\lambda_M/M$ vs. energy. The various curves cross at the position of the mobility edge $E_c/E_\sigma=0.032\pm0.002.$ 
(b): The localization length $\xi$ vs. energy, in the localized regime. It shows an algebraic divergence near the mobility edge, indicated by the red dashed line.}
\label{Fig:Crossing}
\end{figure}

In 3D, the Anderson transition takes place in the regime of strong disorder, so that 
no analytic prediction is available and one has to resort to the numerical calculations in order to compute the position of the mobility edge as well as the localization length below it. We used the transfer matrix method described in~\cite{Delande14}.
To make a long story short, we discretize the configuration space on a (sufficiently dense) 3D rectangular grid and recursively compute the total transmission
of a bar-shaped grid with length $L$ and square transverse section $M\!\times\!M$, with $M\!\ll\!L$.
This system can be viewed as quasi-1D and is thus 
Anderson localized: its total transmission 
decays like $\exp(-2L/\lambda_M)$ where $\lambda_M$ is the quasi-1D localization length in units of the lattice spacing. In practice, the log of the total transmission is a self-averaging quantity
which can be safely computed.
$\lambda_M$ depends on $M,$ on the energy and on the disorder strength. 
Figure~\ref{Fig:Crossing}(a) shows the ratio $\lambda_M/M$ as a function
of energy, for various $M$ values, at a fixed disorder strength $V_0\!=\!E_\sigma.$
At low energy, in the localized regime, $\lambda_M/M$ decreases with $M$ and eventually behaves like 
$\lambda_{\infty}/M$, with $\lambda_{\infty}\!=\!\lim_{M\to\infty} \lambda_M$ the 
3D localization length.
In contrast, at high energy,  $\lambda_M/M$ increases with $M$, a signature of the diffusive regime.
At the mobility edge, $\lambda_M/M$ is a constant $\Lambda_c$ of order unity, meaning that the quasi-1D localization length is comparable to the transverse size of the system, a signature of marginal 3D localization. 
Thus, the mobility edge can be obtained by looking at the point where all curves cross in Fig.~\ref{Fig:Crossing}(a), near $E_c/E_\sigma=0.03.$ In order to pinpoint more accurately
the position of the mobility edge, we use a finite-size scaling analysis~\cite{Delande14,Slevin14} which gives $E_c/E_\sigma=0.032\pm0.002.$ It also makes it possible to compute the 3D localization length
below the mobility edge, shown in Fig.~\ref{Fig:Crossing}(b). In fact, this algebraic  divergence of the localization length near the critical energy shown in Fig~\ref{Fig:Crossing}(b) is a characteristic feature of the localized-delocalized Anderson transition in 3D.
Notice that the diverging localization length appears in units of the correlation length of the disordered potential.
In this regard, since in our model is possible to decrease the correlation length upon increasing $k_0$, it makes feasible the observation of the Anderson transition in a system with finite size. That is, any point in Fig.~\ref{Fig:Crossing}(b) can be realized in our finite system by a choice of sufficiently small $\sigma$. Then, regardless how big $\xi/\sigma$ is, it is always possible to choose such a small correlation length of the effective disordered potential that $\xi$ will be smaller than the system size, i.e. $\xi\ll2\pi$.

\begin{figure}
	\includegraphics[width=0.9\columnwidth]{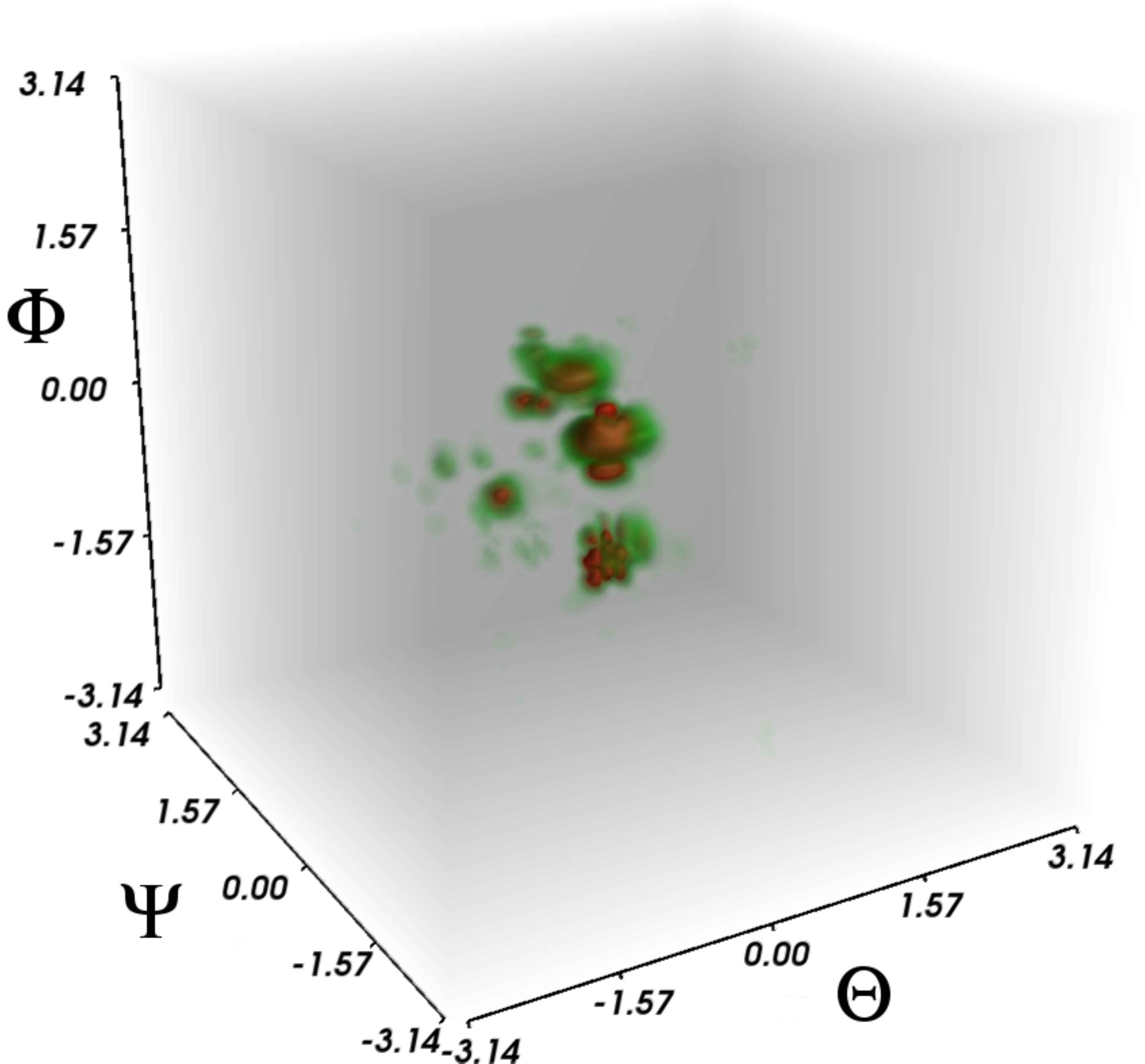}
	\vskip 0.5cm
	\includegraphics[width=0.9\columnwidth]{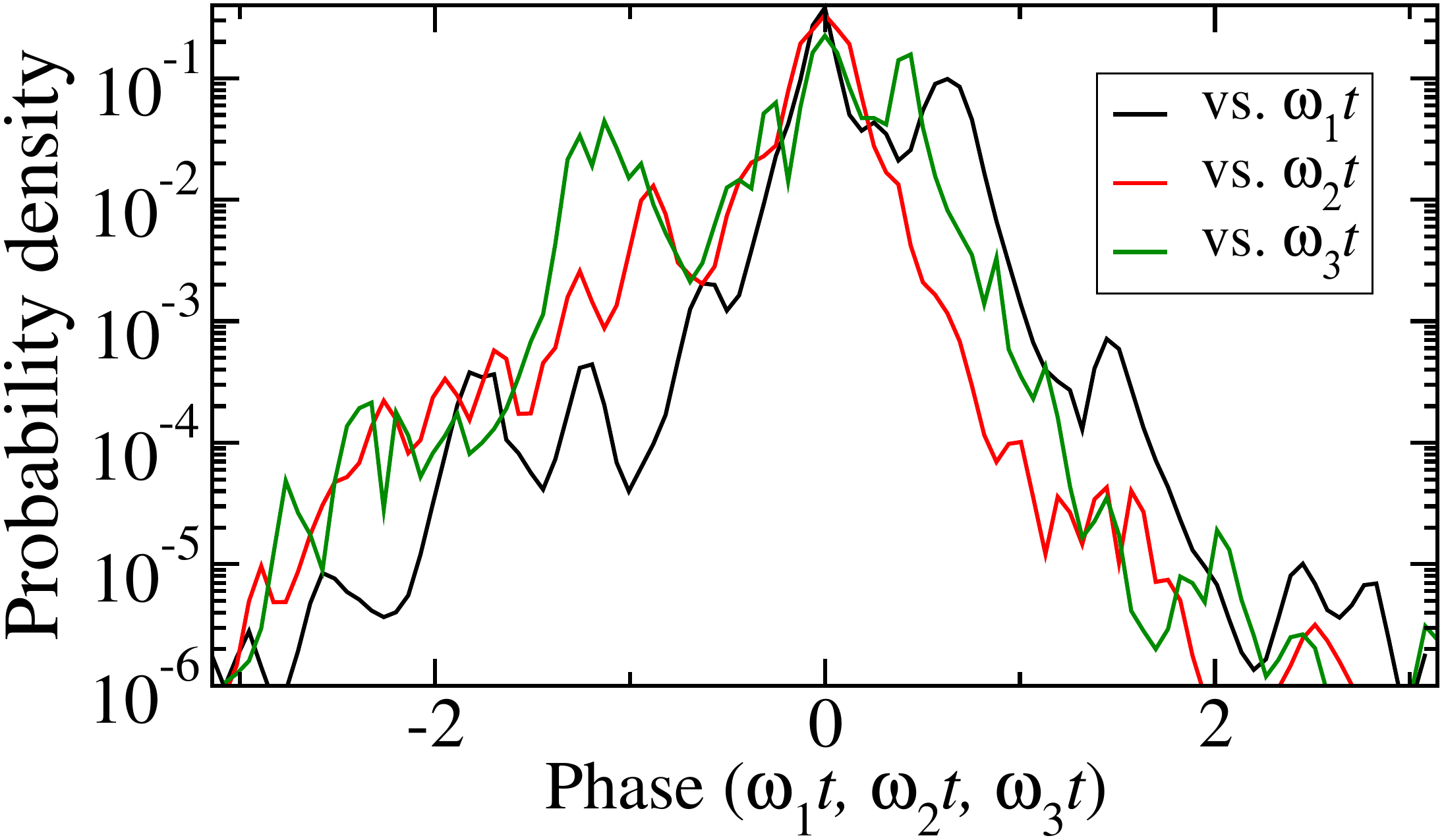}
	\caption{Spatial probability density for a typical localized eigenstate of Hamiltonian (\ref{heff}) with correlation length of the disorder $\sigma=0.1$, for $V_0=E_\sigma$ and energy $E/E_\sigma=-0.049968.$ The upper color 3D plot shows the disordered yet localized character of the state below the mobility edge. Lower plots show how probability densities at a fixed position in $\theta$, $\phi$ or $\psi$ in the laboratory frame (integrated along two remaining directions) evolve in time. The semi-logarithmic scale indicates an approximate exponential localization. In the rotating frame, the localization length $\simeq 0.4=4\sigma$ is in good agreement with the prediction of the transfer matrix calculation, $\xi=4.3\sigma$ in Fig.~\ref{Fig:Crossing}. It implies that in the laboratory frame the localization length in time reads, e.g., $4\sigma/\omega_1$ if the probability density is integrated over $\psi$ and $\phi$.}
	\label{Fig:Density}
\end{figure}

Coming back to the initial time-dependent driven Hamiltonian (\ref{h3d}), observing a temporal dependence with time crystal properties requires to have a periodic motion localized along the 3 directions, that is a stationary eigenstate of the effective Hamiltonian (\ref{heff}) with a localization length much smaller than the spatial period $2\pi.$ By inspecting the results in Fig.\ref{Fig:Crossing}(b), we chose an exemplary value $\sigma=0.1$ (corresponding to $k_0=10\sqrt{2}).$ At energy $E=-0.05E_\sigma,$ the localization length is predicted to be $\xi\approx 4.3\sigma=0.43,$ sufficiently smaller than $2\pi$ to observe good localization properties. We numerically diagonalized the Hamiltonian (\ref{heff}) discretized on a 100x100x100 grid using the JADAMILU package~\cite{Jadamilu} to obtain few eigenstates with energy close to $-0.05E_\sigma.$ The localization properties of a typical eigenstate are shown in Figure~\ref{Fig:Density}. As expected, they display an overall exponential localization with the expected localization length and with the large fluctuations typical of eigenstates. The lower plots show how the probability density for detecting a particle at a fixed point in the configuration space changes as time evolves. It changes periodically with the maximum value roughly 5 orders of magnitude larger than the minimum, i.e. it behaves like the probability in the case of Anderson localization in a space crystal with periodic boundary conditions.
		Other eigenstates, either at slightly different energy and/or for a different disorder realization, have similar localization properties.
		     
\begin{figure}
\includegraphics[width=1.\columnwidth]{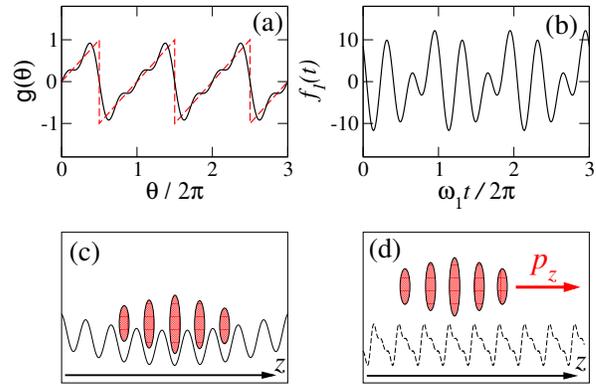}
\caption{Experimental proposal. (a) The red dashed line shows the shape of a sawtooth potential while the black solid line its approximation built with the first three spatial harmonics only. The latter can be created by means of an optical standing wave with wavelength $\lambda$ and its first two harmonics. (b) Temporal modulation function $f_1(t)$ that consists of three harmonics with random phases and with the amplitude $|g_k f^{(1)}_k|=\frac{1}{\sqrt{k_0}}$ for $|k|\le k_0=3$. (c) Schematic plot of the initial stage of the experiment: ultra-cold bosonic atoms are prepared in a strong optical lattice and shallow trapping potentials. We assume that for the amplitude of the lattice potential of the order of $30E_{\mathrm{rec}}$ (atomic recoil energy), slices of the atomic cloud are formed that consist of well defined numbers of particles and do not have mutual phase coherence. The average kinetic energy along the $z$ direction is $\langle p_0^2\rangle/2m\approx2E_{\mathrm{rec}}$. (d) Final stage of the experiment: for a perturbation amplitude $V_0=40E_{\mathrm{rec}}$, atoms accelerated to the average momentum $\langle p\rangle=m\omega_1 \lambda/4\pi$ will fly over the time modulated sawtooth potential and do not spread along $z$ due to the predicted Anderson localization. The predictions are valid provided $\omega_1\ge 400E_{\mathrm{rec}}/\hbar$.
}
\label{experiment}
\end{figure}

Finally, let us analyze a possible experimental realization of Anderson localization in the time domain with ultra-cold atomic gases. We will focus on the simplest version when a single frequency driving is applied and only one spatial degree of freedom is involved. In order to realize a system described by the Hamiltonian $H=p^2/2m+V_0g(2\pi z/L_z)f_1(t)$, which is analogous to (\ref{h3d}), we can use a sawtooth shape periodic potential along the $z$ direction with spatial period $L_z$ \cite{denisov07,salger09}. Initially, an ultra-cold atomic cloud should be prepared in a shallow trap and in the presence of a strong optical lattice along the $z$ axis which has to be periodic with period $L_z$ but can have an arbitrary shape. This creates a series of independent slices of the atomic cloud which consist of well defined numbers of atoms but do not have any mutual phase coherence, see Fig.~\ref{experiment}(c). Next, the initial optical lattice and the shallow trapping potentials are turned off while the sawtooth potential temporally modulated by the $f_1(t)$ function, is turned on. At the same moment, the scattering length of atoms is adjusted to zero by means of a Feshbach resonance and atoms are kicked so that their average momentum along the $z$ direction is $\langle p\rangle=\omega_1mL_z/2\pi$, see Fig.~\ref{experiment}(d). If these conditions are met, the effective Hamiltonian, in the frame moving with the velocity $\omega_1L_z/2\pi$, reads $H_{\mathrm{eff}}=P^2/2m+V_0\sum_kg_kf_{-k}^{(1)}e^{ik2\pi z/L_z}$ and Anderson localization along the $z$ direction can be expected. This requires the localization length $\xi(E)$ corresponding to $E=\langle p_0^2\rangle/2m$ to be smaller than $L_z$ where $\langle p_0^2\rangle$ is the initial dispersion of momenta of atoms along the $z$ axis in the presence of the strong optical lattice potential at the beginning of the experiment. An example of experimental parameters is given in Fig.~\ref{experiment}. It is not necessary to create an exact sawtooth periodic potential. For time modulation $f_1(t)$ consisting of, e.g., three harmonics ($k_0=3$), only the first three spatial harmonics of the sawtooth potential have to be reproduced. The presence of Anderson localization in time will have remarkable signatures in the described experiment. That is, after the turning off the initial optical lattice and trapping potentials, atoms expand slowly in the transverse directions but the width of the slices along the $z$ direction remains smaller than $L_z$ despite the fact that atoms fly over the time modulated sawtooth potential. Although this is a simple setup to implement in the lab, our formulation is not limited to a sawtooth shape for $g(x)$, thus making possible the observation of this phenomenon in a more general setting.  

In conclusion, we have shown that, using a properly disordered, but pseudo-periodic, temporal driving of a 3D system, one can induce a non-trivial Anderson localization in the time domain and the localized-delocalized Anderson transition. This could be observed -- in particular, but not exclusively, using cold atoms -- through the existence of periodically
	evolving localized wavepackets displaying properties similar to those of space crystals with disorder and with periodic boundary conditions but in the time domain.

This work was performed within the Polish-French bilateral POLONIUM Grant 33162XA and the FOCUS action of Faculty of Physics, Astronomy and Applied Computer Science of Jagiellonian University. We thank the authors of the Jadamilu library~\cite{Jadamilu} that we used for large scale diagonalizations. This work was granted access to the HPC resources of TGCC under the allocations 2016-057644 and A0020507644 made by GENCI (Grand Equipement National de Calcul Intensif) and to the HPC resources of MesoPSL financed by the Region Ile de France and the project Equip@Meso (reference ANR-10-EQPX-29-01) of the programme Investissements d'Avenir supervised by the Agence Nationale pour la Recherche. The
work was performed with the support of EU via Horizon2020
FET project QUIC (No. 641122). Support of the National Science Centre, Poland via project No.2016/21/B/ST2/01095 (KS) is acknowledged.

\bibliographystyle{apsrev4-1}

%

\newpage

\section*{Supplemental Material}

We here give the details of the derivation of the effective Hamiltonian, Eq.~(3) in the Letter, within the first order secular approximation and discuss the second order contributions \cite{Buchleitner2002,Lichtenberg_s}.

We consider the Hamiltonian of a particle in the 3D space with periodic boundary conditions with the form
\be
H=\frac{p_\theta^2+p_\psi^2+p_\phi^2}{2}+V_0g(\theta)g(\psi)g(\phi)f_1(t)f_2(t)f_3(t), 
\label{h3ds_s}
\ee
where $V_0$ stands for the amplitude of the perturbation, the angles $\theta$, $\psi$ and $\phi$ denote the position of the particle on the 3D torus and $p_\theta$, $p_\psi$ and $p_\phi$ are 
the conjugate momenta. The time dependent functions are periodic,
\be
f_i(t+2\pi/\omega_i)=f_i(t)=\sum_{k}f_k^{(i)}e^{ik\omega_i t},
\ee
where $f_k^{(i)}=f_{-k}^{(i)*}$ and $f_0^{(i)}=0$. We assume that the ratios of the frequencies $\omega_i$ are irrational numbers.
The function $g(x)=x/\pi$ for $x\in[-\pi,\pi[$, has the following Fourier expansion:
\be
g(x)=\sum_{n}g_ne^{inx},
\ee
where $g_n\!=\!\frac{i(-1)^{n}}{\pi n}$ for $n\ne0$ and $g_0=0$. 

We are interested in the resonant motion when $\theta$, $\psi$ and $\phi$ are changing with frequencies close to $\omega_1$, $\omega_2$ and $\omega_3$, respectively.
Let us perform the canonical transformation to the moving frame, 
\bea
\Theta=\theta-\omega_1 t, &\quad& P_\Theta=p_\theta-\omega_1, \\
\Psi=\psi-\omega_2 t, &\quad&  P_\Psi=p_\psi-\omega_2\\
\Phi=\phi-\omega_3 t, &\quad&  P_\Phi=p_\phi-\omega_3, 
\eea
that results in 
\bea
H&=&\frac{P_\Theta^2+P_\Psi^2+P_\Phi^2}{2}\cr 
&& +V_0\sum_{kmn}\sum_{opr}g_kg_mg_nf_o^{(1)}f_p^{(2)}f_r^{(3)} \cr
&& \times e^{i(k\Theta+m\Psi+n\Phi)}e^{i(k+o)\omega_1 t}e^{i(m+p)\omega_2 t}e^{i(n+r)\omega_3 t},
\label{origh_s}
\eea
where the constant additional term $(\omega^2_1+\omega^2_2+\omega^2_3)/2$ has been omitted.
The new variables $\Theta$, $\Psi$ and $\Phi$ are slowly varying quantities if we choose the conjugate momenta $P_\Theta\approx 0$,  $P_\Psi\approx 0$ and $P_\Phi\approx 0$, i.e. if we focus on the  motion of the particle in the vicinity of a resonant trajectory. Then, the dynamics of the slowly varying variables can be described by an effective Hamiltonian obtained by averaging the original Hamiltonian (\ref{origh_s}) over time,
\bea
H_{\mathrm{eff}}&=&\frac{P_{\Theta}^2+P_{\Psi}^2+P_\Phi^2}{2} \cr &&
+V_0\sum_{kmn}g_kg_mg_nf_{-k}^{(1)}f_{-m}^{(2)}f_{-n}^{(3)}e^{i(k\Theta+m\Psi+n\Phi)}.
\label{effh_s}
\eea
Equation (\ref{effh_s}) is identical to Eq.~(3) in the Letter. In the following we assume that the absolute values of $f_k^{(i)}$ fulfill 
\begin{equation}
|g_kf_{-k}^{(i)}|=\frac{1}{\sqrt{k_0}\pi^{1/4}}e^{-k^2/(2k_0^2)}.
\label{gaussian_s}
\end{equation} 

The effective Hamiltonian (\ref{effh_s}) is the first order secular approximation and it constitutes an accurate description of the resonant dynamics of the particle provided the second order terms can be neglected \cite{Buchleitner2002,Lichtenberg_s}. The latter are proportional to 
\be
\frac{V_0^2}{(k\omega_1+m\omega_2+n\omega_3)^2}\exp\left(-\frac{k^2+m^2+n^2}{4k_0^2}\right), 
\label{smalld_s}
\ee
where $k$, $m$ and $n$ are non-zero integers. Even if the ratios of $\omega_i$ are irrational numbers, small denominators can arise in (\ref{smalld_s}). To avoid it is sufficient to choose:
\be
\omega_1>2k_0(\omega_2+\omega_3).
\label{cond1_s}
\ee 
Then, the exponential function in (\ref{smalld_s}) {\it kills} the second order terms whose denominators are small. 

Thus, when the condition (\ref{cond1_s}) is fulfilled and $V_0^2/\omega_1^2$ goes to zero, the second order contributions become negligible and the effective Hamiltonian (\ref{effh_s}) provides a quantitative description of the resonant behavior of the system. Even if one chooses frequencies $\omega_i$ whose ratios are rational numbers, the second order terms are still negligible if (\ref{cond1_s}) is satisfied, provided one uses $f_i(t)$ functions with vanishing high order Fourier components $f_k^{(i)}\equiv0$ for $|k|>k_0$
(instead of the Gaussian function in Eq.~(\ref{gaussian_s})).

\end{document}